\documentclass[12pt]{article}
\usepackage[cp1251]{inputenc}
\usepackage{epsf}
\usepackage{pazh}
\usepackage[russian]{babel}
\usepackage{morefloats}
\usepackage{graphicx}
\tightenlines

\def\etal{{et~al.}}
\sloppypar

\begin{document}
\baselineskip 21pt

\title{\bf GRB 200415A: magnetar giant flare or short gamma-ray burst?}

\author{\bf \hspace{-1.3cm}\copyright\, 2020  \ \
P. Yu. Minaev\affilmark{1*}, A. S. Pozanenko\affilmark{1}}

\affil{
{\it $^1$ Space Research Institute, Moscow, Russia}}

\vspace{2mm}

\sloppypar \vspace{2mm} \noindent

A detailed analysis of the GBM/Fermi experiment data is carried out to classify GRB 200415A. It is shown that, on the one hand, this event exhibits typical for type I (short) gamma-ray bursts (GRBs) properties, such as duration, variability, and the character of spectral evolution (lag). On the other hand, the localization of the event source on the celestial plane, obtained by the triangulation method (IPN), indicates the nearby ($ D_L $ = 3.5 Mpc) galaxy NGC 253 (Sculptor) as a possible host galaxy for this burst. It introduces significant restrictions on the energetics of the event ($ E_{iso} $ $ \sim $ $ 10^{46} $ erg) and gives an alternative interpretation of GRB 200415A as a giant flare (GF) of a soft gamma repeater (SGR). This interpretation is supported by the atypically hard energy spectrum. In addition, according to the position of the burst on the $ E_{p, i} $ -- $ E_{iso} $ (the position of the maximum in the energy spectrum $ \nu F_{\nu} $ in the source frame depending on the isotropic equivalent of the total energy, emitted in gamma rays) and $ T_{90, i} $ -- $ EH $ (duration in the source frame depending on the combination of parameters $ EH = E_{p, i, 2} ~ E_{iso, 51}^{ ~ -0.4} $) diagrams, GRB 200415A is unambiguously classified as a giant flare of a magnetar, assuming its association with the galaxy NGC 253. In these diagrams, known giant flares in the Galaxy and candidates for giant flares in nearby galaxies form a well-defined group, which includes the GRB 200415A.

\noindent {\bf key words:\/} gamma-ray transients, gamma-ray bursts, GRB, soft gamma repeaters, SGR, magnetars, GRB 200415A, NGC 253, Sculptor galaxy

\noindent

\vfill
\noindent\rule{8cm}{1pt}\\
{$^*$ $<$minaevp@mail.ru$>$}

\clearpage

\section*{INTRODUCTION}
\noindent

Two different types of gamma-ray bursts (GRBs) were discovered in a series of KONUS experiments (Mazets et al., 1981) and then confirmed on larger statistical material in the BATSE/CGRO experiment (Kouveliotou et al., 1993). Type I (short) bursts are characterized by a shorter duration (generally less than 2 s), a harder energy spectrum (a greater portion of high-energy gamma rays), and less pronounced spectral evolution (lag of low-energy gamma rays) compared to type II bursts (Kouveliotou et al., 1993; Norris et al., 2005; Minaev et al., 2010a; Minaev et al., 2012; Minaev et al., 2014). At the same time, the duration and hardness ratio distributions of these two types of gamma-ray bursts, traditionally used for the classification of bursts, overlap significantly, keeping the classification problem actual, especially in the overlap region of the distributions (see, for example, Minaev et al., 2010b; Minaev, Pozanenko, 2017). Correct classification is crucial to investigating the progenitors of gamma-ray bursts.

Type I gamma-ray bursts are associated with the merger of a binary system of neutron stars (Blinnikov et al., 1984; Paczynski, 1986; Meszaros, Rees, 1992), which was recently confirmed by LIGO/Virgo gravitational wave detectors for the GRB 170817A (Abbott et al., 2017a; Abbott et al., 2017b; Pozanenko et al., 2018) and for the GRB 190425A (Abbott et al., 2020; Pozanenko et al., 2020a). Some type I bursts are accompanied by an additional component with a duration of tens of seconds and a softer energy spectrum (compared to the main emission episode) -- the extended emission, which nature has not yet been clarified (Connaughton, 2002; Gehrels et al., 2006; Rosswog, 2007; Metzger et al., 2008; Minaev et al., 2010a; Minaev et al., 2010b; Norris et al., 2010; Barkov, Pozanenko, 2011).

Type II gamma-ray bursts are associated with the core collapse of a massive star (Woosley, 1993; Paczynski, 1998; Meszaros, 2006), some of them are also accompanied by a bright type Ic supernova (Galama et al., 1998; Paczynski, 1998; Cano et al., 2017; Volnova et al., 2017).

Short and hard gamma-ray emission is also characteristic of soft gamma repeaters (SGR) during the giant flares (see, for example, Mazets et al., 1979; Thompson, Duncan, 2001; Frederiks et al., 2007a; Mazets et al., 2008). The light curve of the giant flare consists of a short (fractions of a second), hard and bright main episode, followed by a long (hundreds of seconds) and much weaker extended emission, characterized by strong variability, including quasiperiodicity (Ferochi et al., 1999; Israel et al., 2005). Most SGRs are located in the Galaxy, however, the main short episode of a giant flare can also be registered from nearby galaxies, and its observed properties will be somewhat similar to those of type I gamma-ray bursts, introducing additional difficulties in the classification of gamma-ray transients (see, for example, Pozanenko et al., 2005; Crider, 2006; Popov, Stern, 2006; Mazets et al., 2008). SGRs are probably associated with magnetars -- highly magnetized single neutron stars (B $ \sim $ $ 10^{14} $ G), the nature of their giant flares remains unclear (Duncan, Thompson 1992; Thompson, Duncan, 1995; Kouveliotou et al., 1999).

GRB 200415A was originally classified as a type I gamma-ray burst (Bissaldi et al., 2020), but the localization area of its source on the celestial plane, obtained using the triangulation method, contains the nearby Sculptor galaxy (NGC 253), which indicates a possible connection of this event with a SGR giant flare in this galaxy (Svinkin et al., 2020b). In this work, we carry out a detailed spectral-temporal analysis of this event in the gamma-ray range using data from the GBM/Fermi experiment in order to determine the nature of its source, including using a new method for the classification of gamma-ray bursts based on the correlation of total energy emitted in gamma rays ($E_{iso}$) and spectral hardness ($E_{p,i}$) of bursts, first proposed in (Minaev, Pozanenko, 2020).

\section*{OBSERVATIONS OF GRB 200415A}

A bright burst of gamma rays GRB 200415A with a duration of about 0.2 s and a hard energy spectrum typical for type I gamma-ray bursts was registered on April 15, 2020 at 08:48:06.56 UT by a number of orbital gamma-ray experiments: GBM/Fermi (Bissaldi et al., 2020), LAT/Fermi (Omodei et al., 2020a), Konus-Wind (Frederiks et al., 2020), SPI-ACS/INTEGRAL (Pozanenko et al., 2020b), ASIM (Marisaldi et al., 2020), Mars-Odyssey/HEND (Svinkin et al., 2020b), BAT/Swift (Svinkin et al., 2020b).

The simultaneous registration by a large number of experiments made it possible to construct a sufficiently accurate localization map of the source on the celestial plane using the IPN triangulation method (Svinkin et al., 2020a, Svinkin et al., 2020b). Using the GBM and LAT experiments on board the Fermi observatory, localization maps were independently constructed, found to be consistent with the IPN localization map, but were not so accurate (Omodei et al., 2020b; Bissaldi et al., 2020; Kunzweiler et al., 2020). The area of the IPN localization map is less than 300 sq. arcmin. with a maximum diameter of 27 arcmin.

Inside the localization region the nearby (D$_L$ = 3.5 Mpc) Sculptor galaxy (NGC 253) is placed, which may be the host galaxy of the source of this burst. In this case, this event can be interpreted as a giant flare of the SGR, primarily because of the total energy value of the event, which is insufficient for gamma-ray bursts ($ E_{iso} $ $ \sim $ $ 10^{46} $ erg, Bissaldi et al., 2020; Svinkin et al., 2020a).

The search for optical counterpart of this event was carried out only by the MASTER group a day after the burst inside the IPN localization region, but no reliable candidates were found, the upper limit for the optical source is 18.9 mag. (Lipunov et al., 2020a; Lipunov et al., 2020b).

\section*{GBM/FERMI DATA ANALYSIS}

The source of the initial data for the GBM/Fermi experiment is the public FTP archive (ftp://legacy.gsfc.nasa.gov/fermi/data/). In the data, a gap was found in the time interval (0.0047, 0.0063) s relative to the trigger, which is probably associated with a telemetry overflow. The GBM/Fermi trigger time is used as the zero on the time scale: April 15, 2020 08:48:06.56 UT.

\subsection*{STRUCTURE OF LIGHT CURVE}

The light curves were analyzed using the TTE data of the most illuminated detectors NaI\_00 -- NaI\_03, NaI\_05, BGO\_00 of the GBM/Fermi experiment. The light curve in the energy range (7, 4000) keV is shown in Fig. 1. It consists of two emission episodes -- a short bright main episode with a duration of about 5 ms and a much weaker, slowly decaying tail with a duration of about 15 ms. The duration parameter $ T_{90} $, the time interval for which the detector registers 90\% of the total number of samples (see, for example, Koshut et al., 1996), for GRB 200415A is $ T_{90} $ = 0.124 $ \pm $ 0.005 s, which is typical for both type I (short) gamma-ray bursts and the main peak of SGR giant flares.

The light curve of the main episode shown in the inset in Fig. 1 with a time resolution of 50 $\mu$s, also has a complex structure and consists of several pulses. The minimum variability scale, defined as the minimum time interval during which the energy flux from the source changes by more than 3 standard deviations, is observed during the rising phase of the initial pulse of the main episode (T $ \sim $ -0.0045 s) and is $dT $~$ \sim $ 50 $\mu$s.

\begin{figure}[h]
\includegraphics[width=\textwidth]{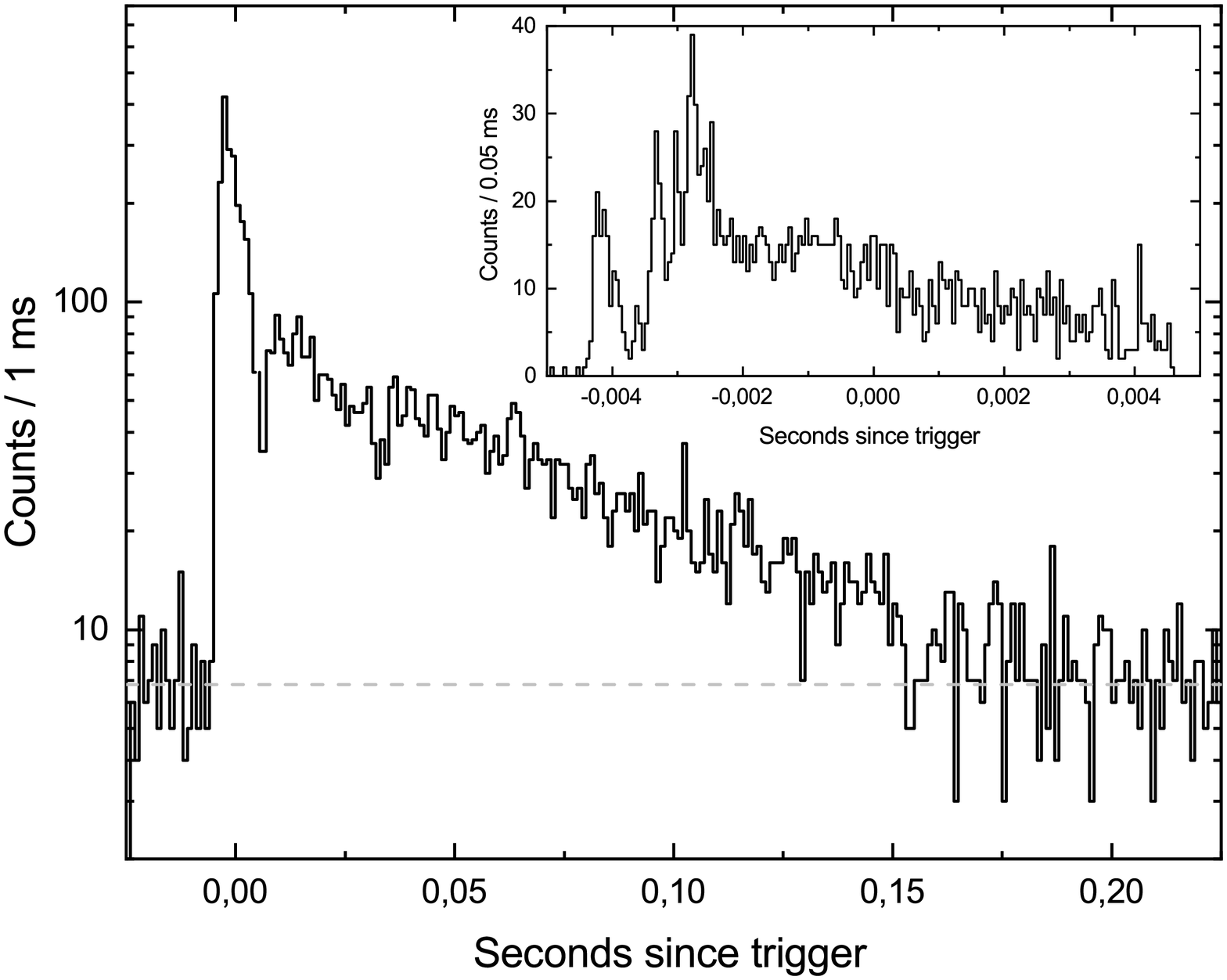}
\\ \textbf{Fig. 1.} \emph{ Light curve of GRB 200415A in the energy range (7, 4000) keV based on GBM/Fermi data with a time resolution of 1 ms, the dashed line shows the background level. The inset shows the light curve of the main episode with a resolution of 50 $\mu$s. The horizontal axis is the time relative to the GBM/Fermi trigger in seconds, the vertical axis is the number of counts in the bin. The gap in the light curve in the interval of (0.0047, 0.0063) s is due to the lack of data.}
\end{figure}

This behavior of the light curve of the main episode is typical for type I gamma-ray bursts. The presence/absence of the fine temporal structure of the light curves of confirmed SGR giant flares is currently not so much known, since: 1) during the giant flares of Galactic sources all space experiments at the time of the main episode are \glqq saturated\grqq; 2) candidates of extragalactic SGR giant flares do not have sufficient count statistics for such studies; 3) there are limitations in the operating modes of some gamma-ray experiments on short time scales (for example, for the Konus-Wind experiment, it is 2 ms, for SPI-ACS/INTEGRAL -- 50 ms).

As noted in the Introduction, GF SGR events are also characterized by long (up to several hundred seconds) extended emission with strong variability, including quasiperiodicity. The relative contribution of the extended emission to the total energetics of the phenomenon varies over a wide range from 1 to 30\% (Mazets et al., 2008). In the GBM/Fermi data for the GRB 200415A, we do not find significant extended emission both in a wide energy range (7, 4000) keV, and in narrower energy channels. With the most conservative estimate of the upper limit on the fluence from extended emission on a scale of 50 s in the range (7, 4000) keV, the contribution of the extended emission component for GRB 200415A is no more than 25\%. Thus, the lack of detection of extended emission does not allow rejecting the hypothesis about the relationship between GRB 200415A and events of the GF SGR class.

\subsection*{SPECTRAL LAG}

As known, GRBs are characterized by spectral evolution, which can be measured as a relative shift (lag) of the light curves in different energy bands. The lag is considered positive if the hard emission is \glqq ahead\grqq~of the soft one, and is determined either by cross-correlation analysis of light curves (see, for example, Minaev et al. 2014), or as a shift in the position of the maximum of the light curve (see, e.g. Hakkila, Preece, 2011). The positive lag characterizes the elementary structures (pulses) of the light curve of gamma-ray bursts, while the negative lag observed in some cases can be a consequence of the superposition effect and arise when analyzing bursts with a complex, multi-pulse light curve structure, since individual pulses have unique properties (Minaev et al., 2014).

In this work, to study the spectral lag, we use the cross-correlation method proposed and described in (Minaev et al., 2014). Light curves are formed in five different energy channels: (7, 70) keV, (70, 200) keV, (200, 400) keV, (400, 900) keV, and (900, 3000) keV. (70, 200) keV is chosen as the reference channel, relative to which the cross-correlation of the remaining channels is carried out. Figure 2 shows the multichannel light curves of the main episode with a temporal resolution of 0.2 ms (left) and the whole event with a resolution of 2 ms (right). The presence of a gap in the time interval (0.0047, 0.0063) s excludes cross-correlation analysis for the entire event, but allows analysis of the main episode, as well as a well-defined initial pulse located near the -0.004 s mark.

\begin{figure}[h]
\includegraphics[width=\textwidth]{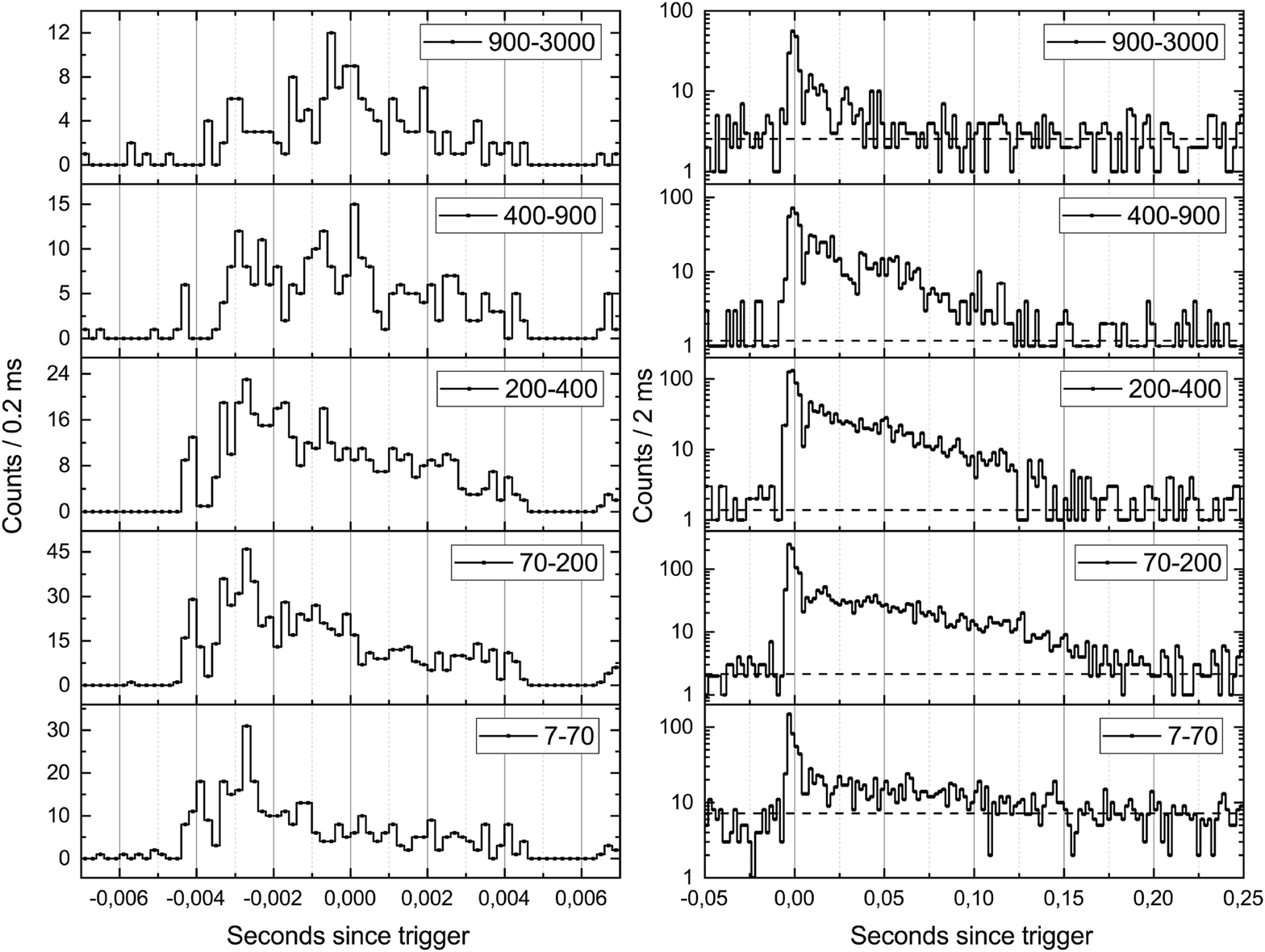}
\\ \textbf{Fig. 2.} \emph{Multichannel light curve of GRB 200415A based on GBM/Fermi data. On the left -- the light curve of the main peak with a time resolution of 0.2 ms, on the right -- the light curve of the whole event with a time resolution of 2 ms, the dashed line shows the background level. The horizontal axis is the time relative to the GBM/Fermi trigger in seconds, the vertical axis is the number of counts in the bin. The boundaries of the energy channels are indicated on the legend. The absence of a signal in the interval (0.0047, 0.0063) s is due to the lack of data.
}
\end{figure}

The results of cross-correlation analysis for the main episode are presented on the left side of Fig. 3. The main episode demonstrates a fast increase of the negative lag, starting from an energy of 400 keV -- the time profile in the hardest range (900, 3000) keV lags behind the softest profile by 1.7 $ \pm $ 0.7 ms. This behavior of the spectral lag can be explained by the superposition effect. Figure 2 shows that the main episode consists of a large number of overlapping pulses, in the energy range (7, 70) keV the maximum of the light curve is near the mark -0.003 s, while in the hard channel (900, 3000) keV -- near zero mark, as the peaks correspond to different pulses of the light curve.

\begin{figure}[h]

\includegraphics[width=\textwidth]{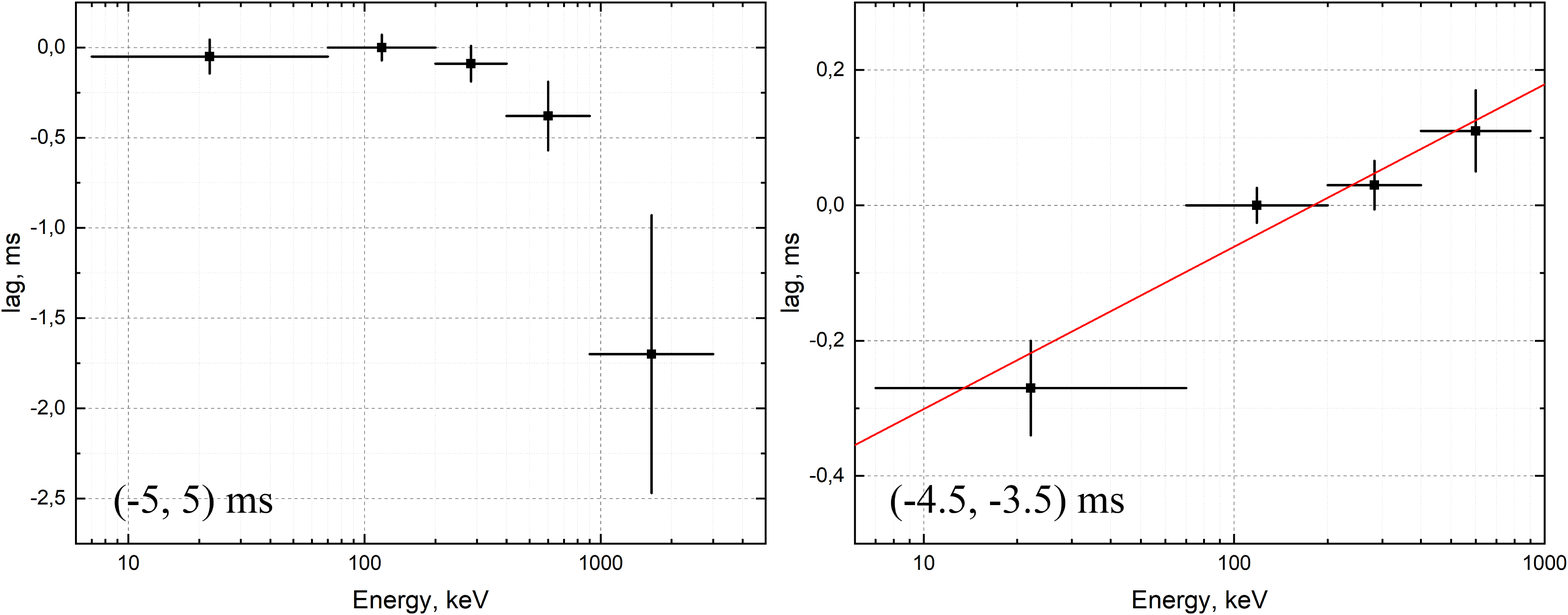}
\\ \textbf{Fig. 3.} \emph{Spectral evolution of GRB 200415A based on GBM/Fermi data. On the left -- for the main episode (time interval (-5, 5) ms relative to the trigger), on the right -- for the initial pulse of the main episode (time interval (-4.5, -3.5) ms relative to the trigger), red line -- approximation by the logarithmic function. The horizontal axis -- the energy in units of keV, the vertical axis -- the spectral lag in units of ms relative to the (70, 200) keV channel.}
\end{figure}

The results of cross-correlation analysis for the initial pulse of the main episode are presented on the right side of Fig. 3 and have a completely different behavior -- a monotonic increase of the spectral lag with increasing energy, which can be described by the logarithmic function $ lag \propto A\log(E) $ with the spectral lag index of $ A = (2.4 \pm 0.9)*10^{-4} $. The change in the position of this pulse, located near the -0.004 s mark, can be easily traced in Fig. 2. This behavior is typical for individual pulses of the light curves of gamma-ray bursts, as shown in Minaev et al. 2014, which may indicate a connection of this event with cosmic gamma-ray bursts. On the other hand, a cross-correlation analysis of giant flares of magnetars has not yet been performed for the same reasons as the estimate of the minimum variability scale (detector overflow, see the previous section).

\subsection*{SPECTRAL ANALYSIS}

To construct and fit the energy spectra, we use the RMfit v4.3.2 software package, specially developed for the analysis of the GBM data of the Fermi observatory (http://fermi.gsfc.nasa.gov/ssc/data/analysis/rmfit/). The spectral analysis method is similar to that proposed in (Gruber et al., 2014). The energy spectra are analyzed using the data of the detectors NaI\_00, NaI\_01, NaI\_05, BGO\_00 of the GBM/Fermi experiment. The energy spectrum of all investigated components of GRB 200415A is not satisfactorily described by both a simple power-law model (PL) and a thermal model (kT), the optimal model is a power-law model with an exponential cutoff (CPL). The results of spectral analysis using this model are presented in Table 1.

\begin{table}[t]

\vspace{6mm} \centering {{\bf Table 1.} Spectral analysis results based on GBM/Fermi data, using power law with exponential cutoff (CPL) model.}\label{tab_spec}

\vspace{5mm}\begin{tabular}{c|c|c|c|c|c} \hline\hline
Interval, $^{1}$ &  $\alpha$  & $E_{p}$,	& Flux, $^{2}$ & $HR_{21}$ $^{3}$ & $HR_{32}$ $^{4}$	\\	

ms 	&  &  keV & $10^{-5}$ erg cm$^{-2}$ s$^{-1}$ & &  \\ \hline

(-6, 150) & 0.05 $\pm$ 0.05 & 976 $\pm$ 44 & 5.51 $\pm$ 0.22 & 5.8 $\pm$ 0.5 & 1.10 $\pm$ 0.04 \\

(-6, 0) & -0.33 $\pm$ 0.06 & 1208 $_{-102}^{+117}$  & 38.9 $\pm$ 2.8 & 3.5 $\pm$ 0.4 & 0.90 $\pm$ 0.06 \\

(0, 150) & 0.21 $\pm$ 0.07 & 929 $\pm$ 47  & 4.4 $\pm$ 0.2 & 7.2 $\pm$ 0.8 & 1.21 $\pm$ 0.05 \\ \hline

(-6, -4) & 0.4 $\pm$ 0.4 & 430 $_{-70}^{+204}$   & 3.4 $\pm$ 0.8 & 5.5 $\pm$ 2.9 & 0.49 $\pm$ 0.14 \\

(-4, -2) & -0.35 $\pm$ 0.09 & 885 $_{-103}^{+123}$   & 42.4 $\pm$ 4.1 & 3.2 $\pm$ 0.5 & 0.73 $\pm$ 0.07 \\

(-2, -0) & -0.3 $\pm$ 0.1 & 1800 $_{-210}^{+250}$   & 75.2 $\pm$ 9.0 & 4.0 $\pm$ 0.8 & 1.12 $\pm$ 0.12 \\

(0, 2) & 0.25 $\pm$ 0.25 & 1690 $_{-212}^{+316}$  & 55.2 $\pm$ 8.0 & 8.9 $\pm$ 4.1 & 1.9 $\pm$ 0.3 \\

(2, 4) & 0.26 $\pm$ 0.27 & 1003 $_{-142}^{+189}$  & 20.0 $\pm$ 3.5 & 7.9 $\pm$ 3.6 & 1.35 $\pm$ 0.24 \\

(8, 14) & 0.63 $_{-0.38}^{+0.48}$ & 1138 $_{-147}^{+226}$  & 14.6 $\pm$ 1.9 & 14.1 $\pm$ 7.7 & 2.1 $\pm$ 0.4 \\

(14, 22) & 1.0 $\pm$ 0.3 & 965 $_{-84}^{+100}$   & 11.3 $\pm$ 1.2 & 24 $\pm$ 15 & 2.3 $\pm$ 0.4 \\

(22, 36) & 0.3 $\pm$ 0.2 & 877 $_{-101}^{+121}$  & 5.2 $\pm$ 0.6 & 8.0 $\pm$ 2.9 & 1.24 $\pm$ 0.15 \\

(36, 50) & 0.42 $\pm$ 0.23 & 885 $_{-91}^{+109}$   & 5.8 $\pm$ 0.7 & 9.8 $\pm$ 3.8 & 1.37 $\pm$ 0.17 \\

(50, 70) & 0.66 $\pm$ 0.32 & 734 $_{-79}^{+99}$   & 3.9 $\pm$ 0.4 & 12.3 $\pm$ 4.7 & 1.31 $\pm$ 0.15 \\

(70, 100) & 0.49 $\pm$ 0.28 & 480 $_{-54}^{+70}$    & 1.43 $\pm$ 0.16 & 7.6 $\pm$ 2.5 & 0.64 $\pm$ 0.08 \\

(100, 150) & 0.47 $\pm$ 0.33 & 377 $_{-50}^{+70}$    & 0.57 $\pm$ 0.07 & 6.1 $\pm$ 2.0 & 0.41 $\pm$ 0.08 \\

\hline

\multicolumn{6}{l}{$^{1}$ -- time interval since GBM/Fermi trigger}\\
\multicolumn{6}{l}{$^{2}$ -- energy flux in range 1 keV -- 10 MeV}\\
\multicolumn{6}{l}{$^{3}$ -- hardness ratio between (50, 300) keV and (15, 50) keV channels}\\
\multicolumn{6}{l}{$^{4}$ -- hardness ratio between (300, 900) keV and (50, 300) keV channels}\\

\end{tabular}
\end{table}

The energy spectrum of the whole event (time interval (-0.006, 0.15) s) has an almost exponential form ($F(E)$ $\propto$ $E^{\alpha}exp\Big[-\frac{(\alpha + 2)E}{E_{p}}\Big]$, $\alpha \simeq 0$)  with the exponential cutoff position $ E_{p} \simeq 1 $ MeV (Table 1). This spectral index value is extremely atypical for short GRBs, for which the value $ \alpha \simeq -0.7 $ is more typical (see, for example, Burgess et al. 2019). This gives rise to doubts about the belonging of this event to the class of short GRBs. Indeed, a similar value of the spectral index was observed in SGR giant flares (see, for example, Frederiks et al. 2007b).

\begin{figure}[h]
\includegraphics[width=\textwidth]{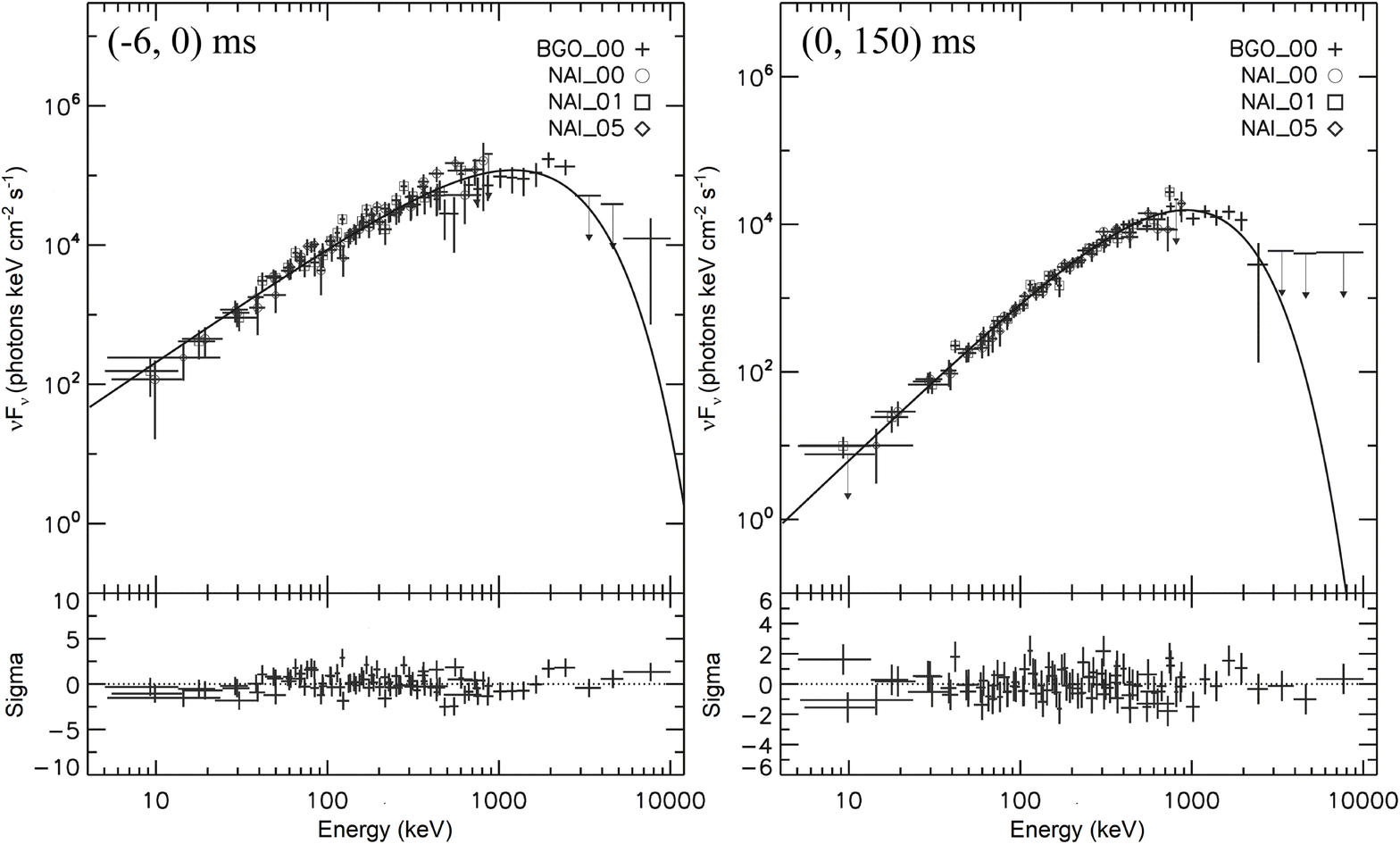}
\\ \textbf{Fig. 4.} \emph{Energy spectrum $ \nu F_{\nu} $ of GRB 200415A based on GBM/Fermi data. On the left -- for the main episode (time interval (-6, 0) ms relative to the trigger), on the right -- for the tail (time interval (0, 150) ms relative to the trigger). The upper panels show the energy spectrum based on the data of the detectors NaI\_00, NaI\_01, NaI\_05, BGO\_00 of the GBM/Fermi experiment. The smooth curve shows the approximation by a power-law model with exponential cut-off (CPL). The lower panels show the deviation of the spectral model from the experimental data, expressed in units of standard deviations.
}
\end{figure}

Assuming that the source of the event is located in the NGC 253 galaxy ($ D_L $ = 3.5 Mpc), the isotropic equivalent of the total energy emitted in the gamma-ray range is $ E_{iso} = (1.26 \pm 0.05)*10^{46} $ erg, which is almost 4 times less than the value for the faintest type I gamma-ray burst GRB 170817A registered at the moment, and is typical for SGR giant flares. On the other hand, the abnormally low value of $ E_{iso} $ for GRB 170817A is connected with observation at a large angle to the jet axis, which, according to various estimates, is about 25 degrees (see, for example, Mooley et al., 2018; Hajela et al., 2019). Then for GRB 200415A, if it is a type I gamma-ray burst with similar emission conditions (energetics, jet opening angle, etc), one can estimate the lower limit on the angle between the observer and the jet axis as $ \sim $ 25 degrees. However, in the case of GRB 170817A, after the main short pulse, a thermal episode with a duration of about 2 s was observed, associated with heating of the shell when the jet reached the surface (Pozanenko et al., 2018; Gottlieb et al., 2018), which is not observed in the case of GRB 200415A.

\begin{figure}
\begin{centering}
\includegraphics[width=0.9\textwidth]{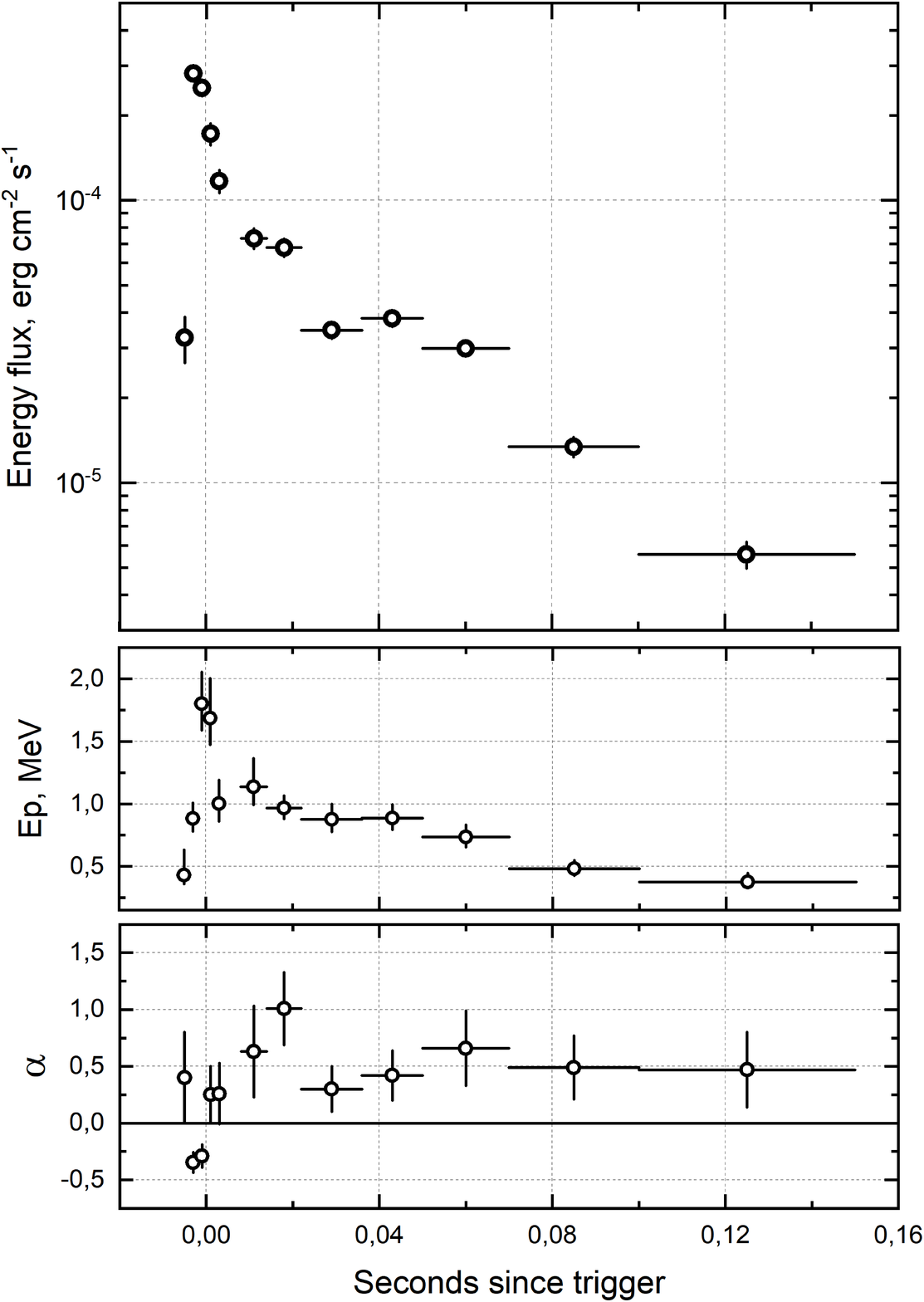}
\end{centering}
\\ \textbf{Fig. 5.} \emph{Spectral evolution of GRB 200415A based on GBM/Fermi data within power law with the exponential cutoff (CPL) model. The upper panel shows the light curve in units of erg cm$ ^{-2} $ s$^{-1}$, the middle panel shows the evolution of the parameter $ E_{p} $ in MeV units, the lower panel -- the evolution of the spectral index $ \alpha $. The horizontal axis is the time in seconds relative to the trigger.
}
\end{figure}

The value of the hardness ratio, calculated as the ratio of the flux in the range (50, 300) keV to the flux in the range (15, 50) keV, expressed in photons and calculated within the optimal model of the energy spectrum, is $ HR_{21} = 5.8 \pm 0.5 $ and along with the previously obtained duration $ T_{90} = 0.124 \pm 0.005 $ s characterizes GRB 200415A as one of the hardest and shortest bursts in the GBM/Fermi experiment (see, for example, Bhat et al., 2016), also confirming its peculiarity.

Spectral analysis of two components of GRB 200415A revealed during the analysis of the light curves: the main episode (time interval (-0.006, 0.0) s) and the tail (time interval (0.0, 0.15) s), confirms their different nature: although the position of the exponential cutoff differs slightly ($ E_{p} \simeq 1 $ MeV), the values of the power indices differ significantly: $ \alpha = -0.33 \pm 0.06 $ for the main episode, $ \alpha = 0.21 \pm 0.07 $ for the tail (Table 1 and Fig. 4).

In addition, the analysis of the spectral evolution is carried out. The light curve is divided into 12 bins with approximately equal signal-to-noise ratio in each bin, in which the energy spectrum is approximated by three models (PL, kT, CPL). The optimal model in all bins is a power-law with exponential cutoff (CPL). The results are presented in Table 1 and Fig. 5. The evolution of the spectrum from soft to hard state is traced within the main episode, reaching the maximum value of $ E_{p} = 1.80_{-0.21}^{+0.25} $ MeV with the index $ \alpha = -0.3 \pm 0.1 $. This behavior is apparently related to the superposition effect -- the light curve of the main episode consists of several pulses with different spectral properties (see the section Spectral lag). After passing the maximum in the light curve, the energy spectrum begins to evolve from hard to soft state (the second component of the light curve -- the tail), which is observed as a shift of the cutoff position down to $ E_{p} \sim 400 $ keV with an almost negligible change in spectral index value, $ \alpha \sim 0.5 $.

Thus, the results of the spectral analysis of GRB 200415A confirm the complex structure of this event and indicate its connection with the class of SGR giant flares.

\subsection*{$E_{p,i}$ -- $E_{iso}$ CORRELATION}

\begin{figure}
\includegraphics[width=\textwidth]{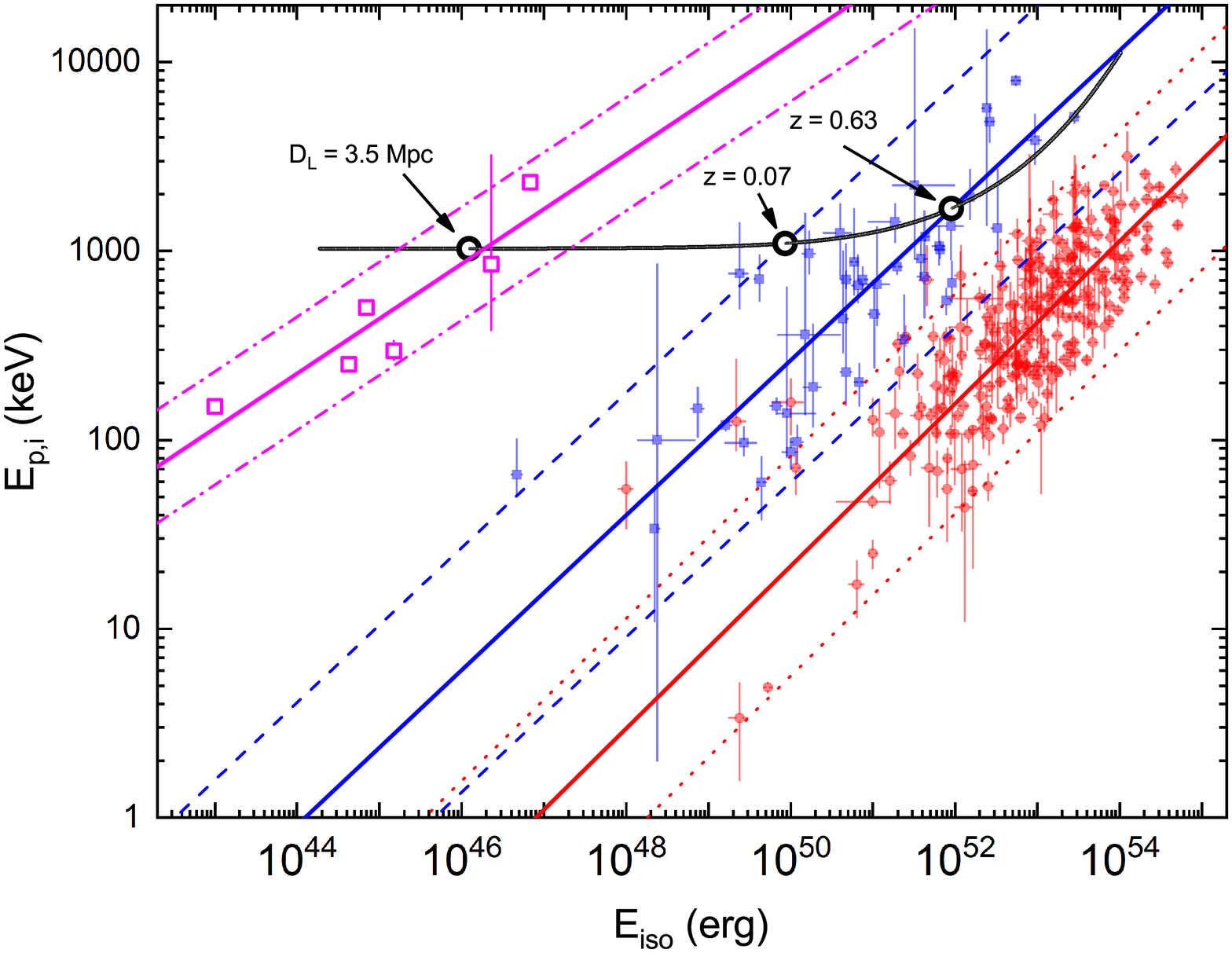}
\\ \textbf{Fig. 6.} \emph{ $ E_{p,i} $ -- $ E_{iso} $ correlation of for type I (blue squares), type II (red circles), and SGR giant flares (pink open squares) with the corresponding approximation results, including 2$\sigma_{cor}$ correlation regions, shown by corresponding colors. The trajectory (dependence on redshift) for GRB 200415A is shown by a smooth black curve. The open black circles correspond to the burst position under the assumption of association with the galaxy NGC 253 (D$_L $ = 3.5 Mpc) and the points of intersection of the trajectory with the upper boundary 2$\sigma_{cor} $ of the correlation region ($z$ = 0.07) and the correlation trend line ($z$ = 0.63) for type I bursts.
}
\end{figure}

It was shown in (Minaev, Pozanenko, 2020) that the correlation between the isotropic equivalent of the total energy emitted in the gamma-ray range, $ E_{iso} $ and the position of the maximum in the energy spectrum $ \nu F_{\nu} $ in the source frame, $ E_{p} $ (formula 1), can be effectively used to classify gamma-ray bursts. This is facilitated by the observational fact that this correlation for various types of gamma-ray bursts is described by a power law with a single index of $ a \simeq 0.4 $, while the correlation region of type I gamma-ray bursts is above the correlation region of type II bursts.

\begin{equation}
    \lg\Big(\frac{E_{p,i}}{100~{keV}}\Big) = a\lg\Big(\frac{E_{iso}}{10^{51}~{erg}}\Big) + b.
	\label{eq:amati}
\end{equation}

However, the parameters $ E_{iso} $ and $ E_{p,i} $ can be calculated only for a source with known distance, which was not determined in the case of GRB 200415A (except for the possible association with the galaxy NGC 253). On the other hand, one can analyze the position of GRB 200415A on the $ E_{p,i} $ -- $ E_{iso} $ diagram depending on the redshift of its source to find out if the trajectory crosses the correlation region for both types of gamma-ray bursts, or intersects only the region of type I bursts. In the latter case, the position of the points of intersection of the trajectory with the region of correlation for type I gamma-ray bursts will make it possible to estimate the distance to the source.

We use a sample of 320 gamma-ray bursts with a known redshift, as well as the results of the analysis of the $ E_{p,i} $ -- $ E_{iso} $ correlation for this sample, published in (Minaev, Pozanenko, 2020). In addition, we include in the analysis six known giant flares of magnetars according to (Mazets et al., 2008). The corresponding $ E_{p,i} $ -- $ E_{iso} $ diagram is shown in Fig.6. It follows from Fig. 6 that giant SGR flares are located separately on the diagram, in the upper left corner as faint but spectrally hard events. Thus, the $ E_{p,i} $ -- $ E_{iso} $ correlation can be used not only to classify gamma-ray bursts, but also to separate type I gamma-ray bursts from SGR giant flares.

Special attention should be paid to the fact that the considered giant flares themselves obey the $ E_{p,i} $ -- $ E_{iso} $ correlation: Spearman correlation coefficient and the corresponding value of the probability of random correlation are $ \rho $ = 0.94 and $ P_{\rho} $ = $ 4.8*10^{-3}$, and while including the GRB 200415A in the sample -- $ \rho $ = 0.93 and $ P_{\rho} $ = $ 2.5*10 ^{-3} $. When approximating the correlation, we obtain the values of the parameters $ a $ = 0.29 $ \pm $ 0.05, $ b $ = 2.4 $ \pm $ 0.3, i.e. the power law index for SGR giant flares within 2$\sigma $ is the same as for gamma-ray bursts, probably indicating a similar emission mechanism. The approximation results are shown in Fig. 6.

The GRB 200415A trajectory, calculated for the spectrum of the whole event in the interval (-0.006, 0.15) s, intersects only the region of SGR giant flares and type I gamma-ray bursts, excluding the connection of this event with type II gamma-ray bursts. If we assume that the source of the burst is indeed in the galaxy NGC 253, then the position of this event on the diagram allows us to unambiguously classify it as a giant flare of a magnetar. On the other hand, if the event source is located in another, more distant galaxy and the event is a type I gamma-ray burst, then the lower limit ($ z $ = 0.07) and the most probable redshift value ($ z $ = 0.63) can be estimated as the intersection points of the trajectory with an upper bound 2$\sigma $ of the correlation region and an approximation curve for type I bursts, respectively.

\subsection*{$T_{90,i}$ -- $EH$ DIAGRAM}

To solve the problem of classification of gamma-ray bursts, another method was proposed in (Minaev, Pozanenko, 2020), which uses, in addition to the $ E_{p,i} $ -- $ E_{iso} $ correlation features, the bimodality of the distribution of gamma-ray bursts in duration in source frame  $ T_{90,i}$. For this purpose, the $ EH $ parameter (Formula 2) was introduced, which characterizes the position of the gamma-ray burst on the $ E_{p,i} $ -- $ E_{iso} $ diagram.

\begin{equation}
    EH = \frac{(E_{p,i} / 100~{keV})}{ (E_{iso} / 10^{51}~{erg})^{~0.4}}.
	\label{eq:EH}
\end{equation}

\begin{figure}
\includegraphics[width=0.95\textwidth]{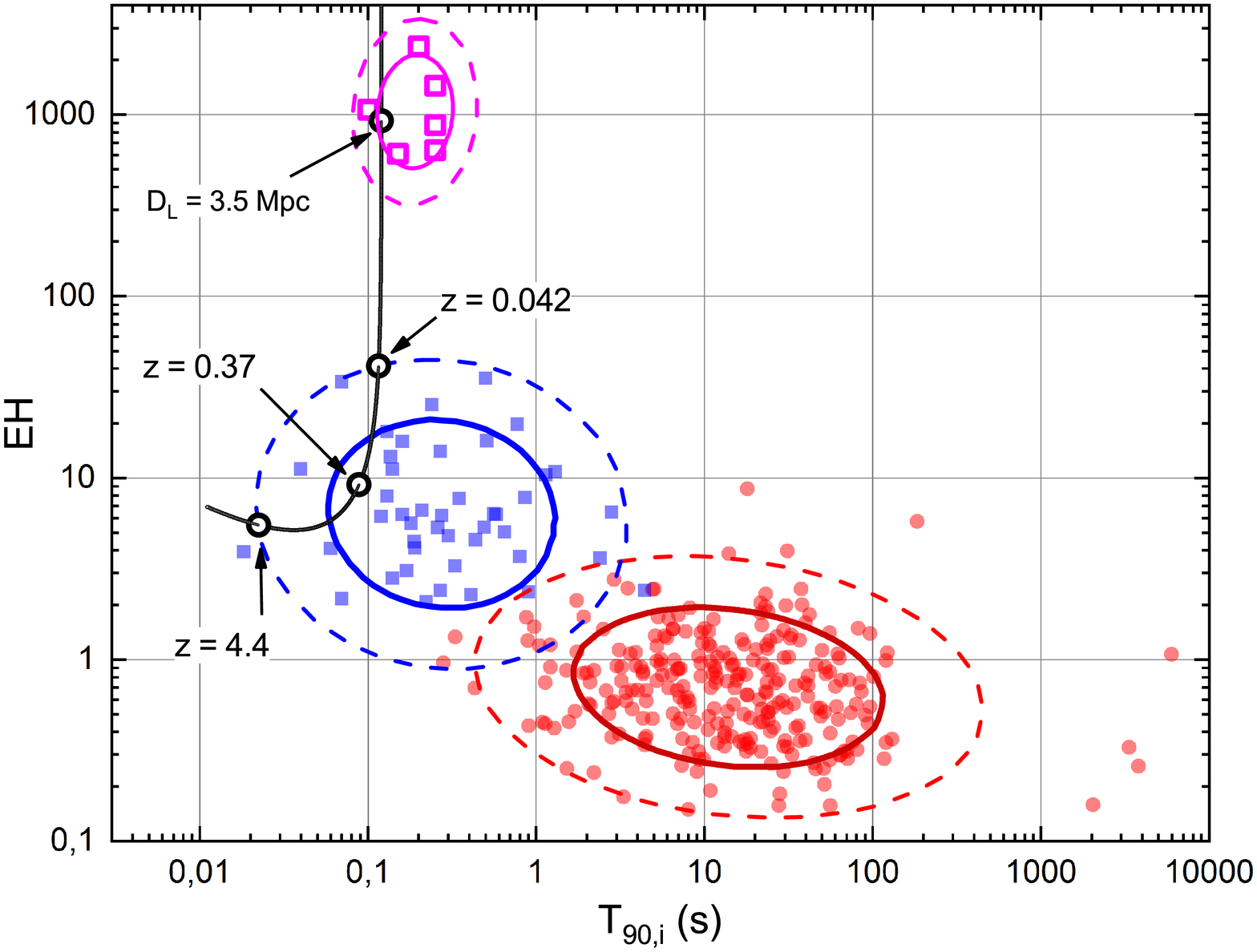}
\\ \textbf{Fig. 7.} \emph{The $ T_{90,i} $ -- $ EH $ diagram for type I (blue squares), type II (red circles) and SGRs giant flares (pink open squares) with corresponding cluster analysis results, 1$\sigma_{cor} $ and 2$\sigma_{cor} $ cluster regions are shown by bold solid and thin dashed curves of the corresponding colors. The trajectory (dependence on redshift) for GRB 200415A is shown by a smooth black curve. Unfilled black circles correspond to the burst position under the assumption of association with the galaxy NGC 253 (D$_L $ = 3.5 Mpc), to the points of intersection of the trajectory with the 2$\sigma_{cor} $ boundaries of the region of the cluster of type I bursts ($z$ = 0.042 and $z$ = 4.4) and the trajectory point closest to the center of the cluster of type I bursts ($z$ = 0.37).
}
\end{figure}

Figure 7 shows the $ T_{90,i} $ -- $ EH $ diagram for 320 gamma-ray bursts from (Minaev, Pozanenko, 2020) and 6 giant SGR flares from (Mazets et al., 2008). Type I gamma-ray bursts, in comparison with type II gamma-ray bursts, have a harder spectrum (in terms of $ E_{p,i} $ values) with a lower value of total energy $ E_{iso} $ and, as a consequence, a larger value of the parameter $ EH$, and also have a shorter duration $ T_{90,i}$. SGR giant flares have the same duration as type I gamma-ray bursts, but much lower energetics with similar spectral hardness, which results in extremely high values of the parameter $ EH$. Thus, the $ T_{90,i} $ -- $ EH$ diagram can also be used not only to classify gamma-ray bursts, but also to separate giant SGR flares from type I gamma-ray bursts.

For three samples of events (type I and II gamma-ray bursts, SGR giant flares), the corresponding clusters on the diagram were approximated by three Gaussians using the expectation-maximization (EM) algorithm -- an iterative method for finding estimates of the maximum likelihood of parameters of probabilistic models depending on several hidden variables. The E-step (expectation) calculates the expected value of the likelihood function, while the hidden variables are treated as observables. In the M-step (maximization), a maximum likelihood estimate is calculated, thus increasing the expected likelihood calculated in the E-step. This value is then used for the E-step in the next iteration. The regions corresponding to 1$\sigma $ and 2$\sigma $ are shown in Fig. 7.

The trajectory of GRB 200415A in the $ T_{90,i} $ -- $ EH $ diagram, depending on the redshift of the source, crosses the regions of SGR giant flares and type I gamma-ray bursts. Assuming that the source is located in the NGC 253 galaxy, the event is unambiguously classified as a SGR giant flare. Otherwise, assuming that GRB 200415A is a type I gamma-ray burst, one can estimate the lower ($ z $ = 0.042) and upper ($ z $ = 4.4) redshift limits as the intersection points of its trajectory with 2$\sigma $ region of the cluster of type I gamma-ray bursts, as well as the most probable value ($ z $ = 0.37) at the point of the trajectory closest to the cluster center.

\section*{CONCLUSIONS}

The work is devoted to determination the nature of the source of GRB 200415A. An analysis of this event in the gamma-ray range was carried out using the data of the GBM/Fermi experiment for the purpose.

Analysis of the light curve revealed the presence of two emission components: a bright and short main episode with a duration of about 5 ms, and a fainter and more prolonged tail with a duration of about 15 ms. The main episode also has a complex structure and consists of several pulses. This behavior of the light curve is typical for type I gamma-ray bursts, while the fine structure of the light curves of known giant flares is poorly determined. Long (hundreds of seconds) extended emission, typical for giant flares of magnetars, was not detected for GRB 200415A. However, the obtained upper limit on the relative flux from extended emission does not exclude the association of the burst with a giant flare of the magnetar. Thus, the features of the light curve do not allow one to make an unambiguous conclusion about the nature of GRB 200415A.

Using cross-correlation analysis, the spectral evolution of the main episode GRB200415A and its well-isolated initial pulse was investigated. It is shown that the dependence of the spectral lag on energy for the initial pulse obeys a logarithmic law with a positive spectral lag index (hard emission is registered earlier than the soft one). At the same time, this dependence for the number of pulses composing the main episode has a more complex form, which can be explained by the effect of superposition of pulses with different properties. The revealed features of the spectral evolution are typical of gamma-ray bursts; however, as in the case of the fine structure of the light curve, they are poorly studied for giant flares of magnetars, which also does not allow making an unambiguous conclusion about the nature of GRB 200415A.

Spectral analysis, carried out both for the time interval covering the entire burst and for its individual components, demonstrated features that are not typical for gamma-ray bursts. Although the energy spectra of all studied components are well described by the power-law model with exponential cutoff (CPL), the power law index has a peculiar, close to zero value for the spectrum of the whole burst, and positive ($ \alpha $ = 0.21 $ \pm $ 0.07) -- for the second emission component. At the same time, some known giant flares of magnetars had similar features. Thus, the results of the spectral analysis do point to an association of GRB 200415A with SGR giant flares.

If GRB 200415A was a short gamma-ray burst (Type I) in the NGC 253 galaxy, then one would expect the registration of the thermal component, similar to its registration in the case of GRB 170817A, where the source was located at a distance of 40 Mpc. This is further evidence in favor of classifying GRB 200415A as a giant flare from SGR (under assumption about its association with the NGC 253 galaxy).

The position of GRB 200415A in the $ E_{p,i} $ -- $ E_{iso} $ and $ T_{90,i} $ -- $ EH $ diagrams is investigated, and it is shown if the burst source is indeed located in the NGC 253 galaxy, which is indicated by the IPN localization of the source on the celestial plane, then it is unambiguously classified as a SGR giant flare, having duration typical for type I gamma-ray bursts ($ T_{90,i} $ = 0.12 s) and the position of the spectrum maximum ($ E_{p,i} $ $ \sim $ 1 MeV), but with very low total energy emitted in the gamma-ray range ($ E_{iso} $ $ \sim $ $ 10^{46} $ erg).

Thus, we classify this burst as a giant flare of a magnetar from the NGC 253 galaxy. The possibility of observing such an outburst from the NGC 253 galaxy was previously indicated in the work of Popov and Stern (2006).

The known giant SGR flares form a well-distinguished group in the $ E_{p,i} $ -- $ E_{iso} $ and $ T_{90,i} $ -- $ EH $ diagrams, similarly to the groups of long and short GRBs. This allows not only to classify the sources of gamma-ray bursts and to distinguish events of the giant flares SGR class, but also to assume the same emission mechanism of SGR giant flares and gamma-ray bursts.


\end{document}